# Fast Charging Limits of Ideally Stable Metal Anodes in Liquid Electrolytes

*Bingyuan Ma[1] and Peng Bai[1,2,*]*

[1] Department of Energy, Environmental & Chemical Engineering, Washington University in St. Louis, St. Louis, MO 63130, United States of America

[2] Institute of Materials Science and Engineering, Washington University in St. Louis, St. Louis, MO 63130, United States of America

*Correspondence to: pbai@wustl.edu

**Abstract:**

Next-generation high-energy-density batteries require ideally stable metal anodes, for which smooth metal deposits during battery recharging are considered a sign of interfacial stability that can ensure high efficiency and long cycle life. With the recent successes, whether the absolute morphological stability guarantees absolute electrochemical stability and safety emerges as a critical question to be investigated in systematic experiments under practical conditions. Here, we use the ideally stable ingot-type sodium metal anode as a model system to identify the fast-charging limits, i.e. highest safe current density, of metal anodes. Our results show that metal penetration can still occur at relatively low current densities, but the overpotentials at the penetration depend on the pore sizes of the separators and surprisingly follow a simple mathematical model we developed as the Young-Laplace overpotential. Our study suggests that the success of stable metal batteries with even the ideally smooth metal anode requires the holistic design of the electrolyte, separator, and metal anodes to ensure the penetration-free operation.

## 1. Introduction

Alkali metal anodes, especially the lithium (Li) metal anode, hold the promise to significantly increase the energy density of current batteries.[1–4] Achieving the ideal interfacial morphological stability of metal anodes during battery recharge, i.e. smooth metal plating, is a primary focus of current research and development of efficient and safe metal anodes.[5,6] In most cases, however, the formation and evolution of the heterogeneous solid electrolyte interphase (SEI)[7] undermine the interfacial stability and lead to rough and porous deposits.[8–10] Even in the high-performance Li metal anodes reported so far, voids, crevices, and nodules are still visible in both optical and electron microscopy observations.[11–18] Ideally smooth and stable Li metal anodes are still under development.

Surprisingly, an ideally smooth, non-porous, ingot-type sodium (Na) metal anode was recently observed in glyme-based liquid electrolytes with strictly controlled moisture concentration.[19] *Operando* observations revealed the high reversibility of the smooth deposits without whiskers, gas bubbles, or disconnected metal structures that were usually observed in other studies.[20–22] The electrodeposition process self-regulate to form a monolithic piece via the classic step movement mechanism. The ideally stable interface indeed enabled stable cycling of anode-free Na metal anodes at high capacities in practical sandwich cells, without resorting to any additional material engineering approaches.[23–26] The special property and superior performance allow us to study a question that all other metal anodes will face: could the ideal morphological and interfacial stabilities guarantee absolute safety?

Here, we use Na as a model system to identify the safety issues the ideally stable metal anodes may encounter during fast charging. Chronopotentiometry experiments reveal the existence of a critical current density that depends on the structural properties of the separator. This critical current is less than 9 % of the corresponding intrinsic limiting current density, even if the ideal morphological stability is in place. Further analyses validate that the observed electrochemical limits can be accurately captured by a Young-Laplace overpotential, based on which a holistic design and management strategy is established. Detailed discussions of the pore selection and in-pore transport dynamics highlight the importance of uniform current distribution at the heterogeneous separator|electrode interface.

## 2. Results

### 2.1 Apparent penetration current densities

While it is widely believed that fast charging can easily trigger metal penetration and internal shorts, quantitative understandings are still missing. Establishing clear mathematical correlations is an outstanding challenging. Chronopotentiometry experiments were conducted in cells with the layered configuration of Cu|separator|washer|Na, as illustrated in Fig. 1a, where the washer defines the reaction area for the accurate determination of current densities. Commercial separators with different (average) pore sizes were used for systematic investigation and cross-validation of the electrochemical dynamics. Details of these separators can be found in Table S1. 1 M $NaPF_6$ in diglyme was used as the electrolyte to achieve the ideally stable Na anodes.[19] For each cell, a constant current density was applied to induce the continuous accumulation of deposits on the anode side. Depending on the current density, two typical transient voltage curves can be observed. As shown in Fig. S1, the voltage curve that remains stable until a divergence due to the depletion of the counter electrode indicates that there was no metal penetration. On the other hand, sudden drops in the voltage curves are reliable signs of metal penetration through the separator.[27] The characteristic capacities taken at the voltage divergence or sudden drop were plotted in Fig. 1b. with respect to the corresponding current densities. A clear critical current density can be identified for each type of separator, as indicated by the vertical dashed lines in Fig. 1b. For current densities lower than the critical value, complete depletion of the counter Na electrode was confirmed after disassembly of the cell, where the mirror-flat smooth deposits[19] (Fig. 1c) are consistent with the stable voltage responses. Higher than this value, penetration-induced sudden voltage drops[27,28] occurred at areal capacities as lower as just 0.005 mAh cm$^{-2}$ for the two commercial separators. Unlike the penetration process by porous Li metal anodes,[27] our Na metal anode, at the onset of metal penetration, was ideally smooth (Fig. 1d), which clearly alerts that the ideal smoothness, or interfacial morphological stability, cannot exclude the possibility of metal penetration through the porous separator, at least for the available commercial separators. To better understand the physics, anodic aluminum oxide (AAO) separators with different pore sizes were adopted for control experiments. The results revealed an inverse relationship between the critical current density and the average pore size of the separator: separators with smaller average pore sizes tend to enable higher critical current density and higher critical areal capacity.

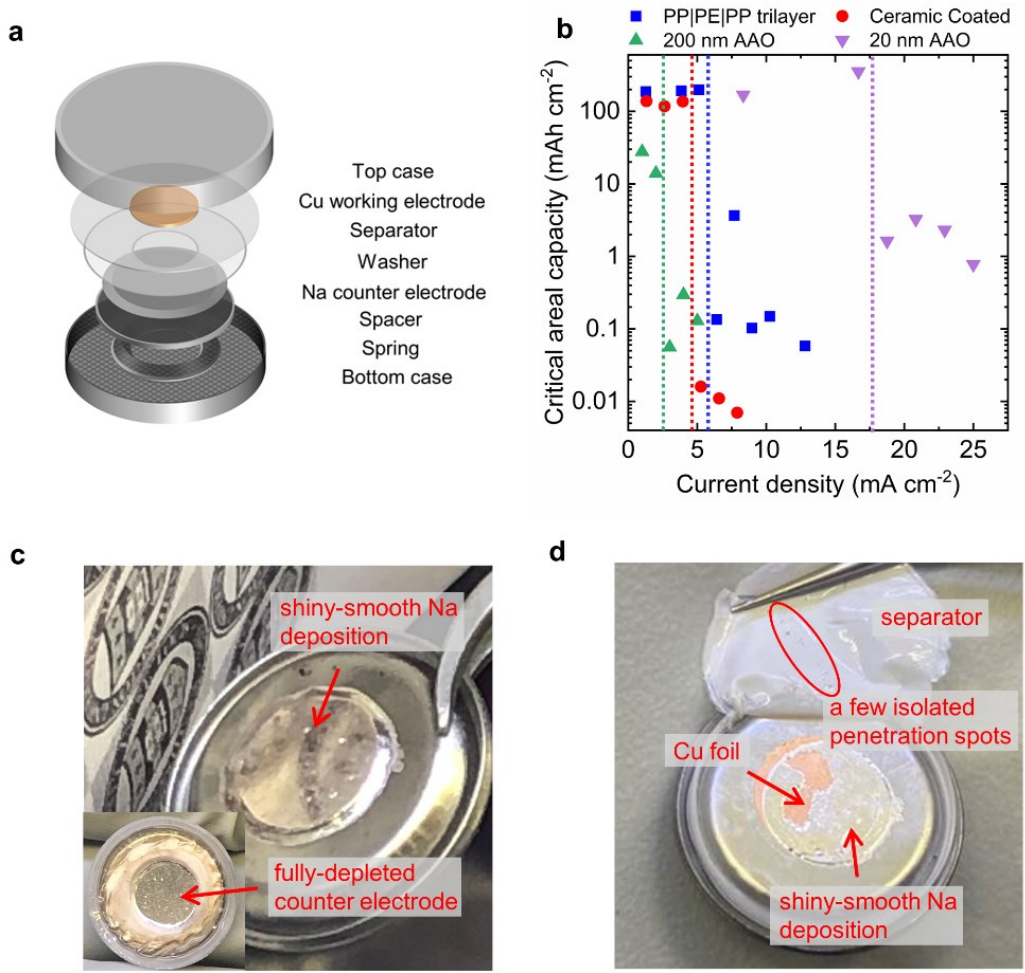

**Fig. 1 Critical apparent current densities of the ideally stable Na metal anode.** (a) Cell configuration. (b) Summary of critical capacities at sudden voltage changes under different current densities. Dashed lines indicate the critical current densities. (c, d) Post-mortem digital images of the deposits from tests at a current density (c) lower or (d) higher than the critical current density.

As indicated by the black spots in the post-mortem images presented in Fig. 1d and those available in the literature,[28] metal penetration always occurs locally through only a few naturally selected pores.[29–31] This observation suggests that the onset of penetration may be attributed to the dynamics at the entrance of just a few favorable pores. A simple mathematical model developed around the most favorable pores, rather than over the entire separator|electrode interface, may capture the key physics and predict the critical limits.

## 2.2 The critical penetration overpotential

The first question regarding the onset of penetration is whether or not Na metal can easily enter the pores. As can be seen in Video S1, a small piece of molten Na can hardly wet the surface of the AAO separator. When the AAO is tilted, the molten Na bead can freely roll down the slope (Video S1). To the best of our knowledge, there is no evidence advising the easy wetting of separators by sodium metal at the room-temperature. It is therefore appropriate to assume that Na metal filaments that may penetrate the separator would form a dewetting contact angle ($\theta$) against the pore wall of the separator, which will be close to 180°, yielding $\cos \theta \approx 1$. As such, an external driving force is required to counteract the dewetting capillary force and push Na into the pores, especially at the very beginning of metal penetration.

Fig. 2a demonstrates the schematic understanding of this process with the key parameters. At the critical moment that the flat Na metal interface is being pushed to deform into the pore of the separator, as there is no wetting capillary force to "pull" the metal into the pores, the driving force should mainly come from the stress in the Na metal electrode (substrate), induced by the addition of new deposits of volume $dV$, after an infinitesimal time, $dt$. We can assume that the internal stress of the Na anode would squeeze Na into $n_p$ pores similar to the most favorable pore drawn in Fig. 2a. The contact circumference between Na and the pore wall will sweep a distance of $dl$ to occupy a total pore volume of the same $dV$, after which the internal stress will get relaxed. Once fully within the pore, the dome-shaped Na interface is relatively closer to the counter electrode, and therefore will induce the localization of the incoming Na$^+$ flux,[27,28] to achieve an increasingly faster reaction. The dynamics within this naturally selected pore will eventually become transport-limited, due to the highly focused flux. As we shall discuss in later sections, a transport-limited Na dendrite will be triggered to chase the retreating concentration front (Fig. 2a rightmost), consistent with the case of Li dendrite growths. Note that any possible internal stress due to the addition of further Na deposits will not dominate the dynamics due to higher local current densities and higher local overpotentials. This critical mechano-electrochemical process at the onset of penetration can be modeled by examining the collective charge balance and force balance within the $n_p$ pores.

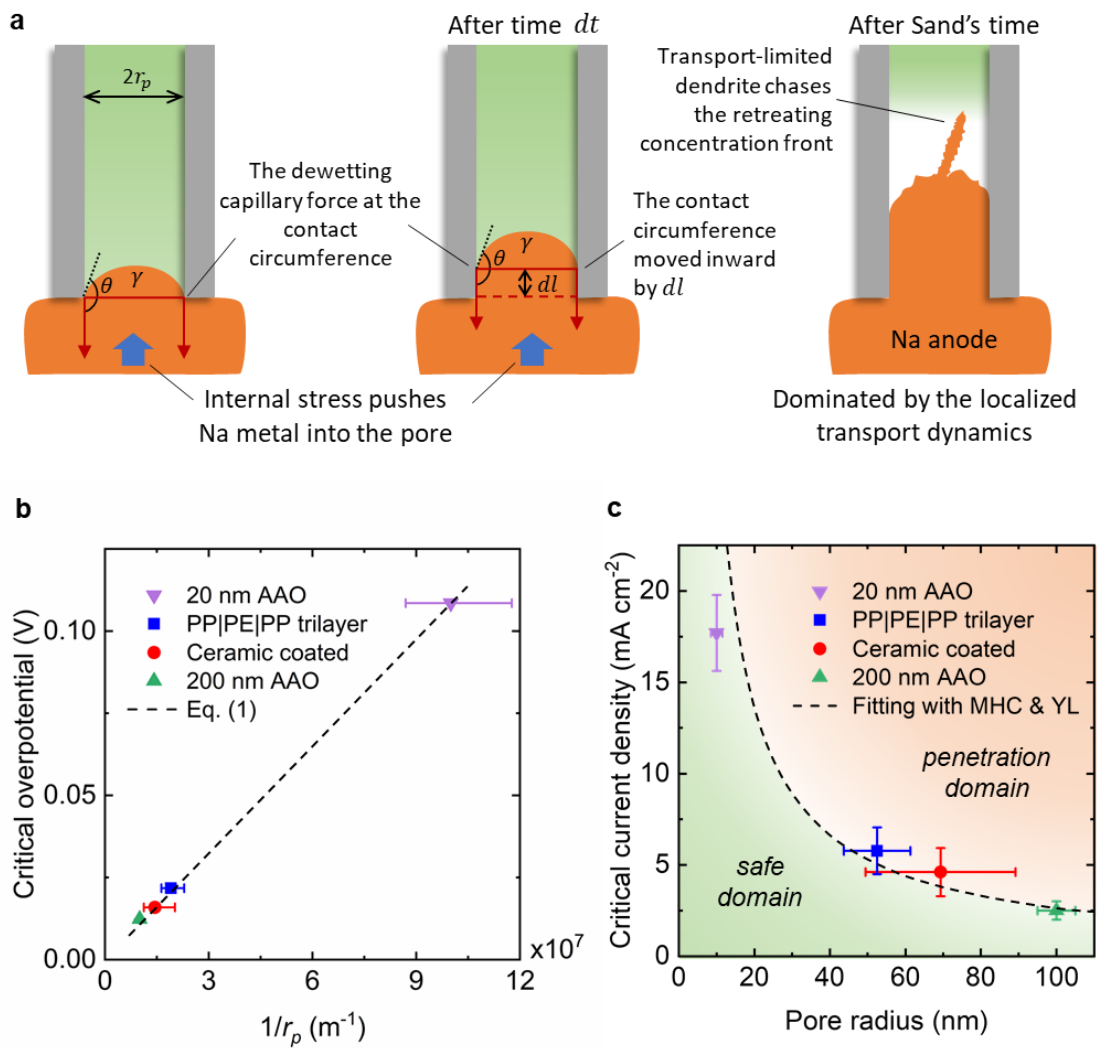

**Fig. 2 Young-Laplace overpoential at the onset of metal penetration.** (a) Schematic understandings of the mechanism at the critical moments of metal penetration and dendrite initiation. (b) Plot of reaction overpotential versus $1/r_p$, yielding the surface energy of Na metal via Eq (1). (c) Determination of the safe-penetration boundary in terms of current densities according to the separator pore sizes. The critical overpotentials were taken from Fig. S1 after being compensated by the I-R drop. $r_p$ is the pore radius with the error bar of the standard deviation. The critical current densities were taken as the average with the range between the measured highest safe value and the lowest penetration-inducing value.

According to the Faraday law, we can establish the charge balance as $dQ = Idt = Fn_p \pi r_p^2 dl/V_m$. Here, $I$ is the total current, $F$ the Faraday constant, $r_p$ the pore radius, and $V_m$ the molar volume of Na metal. To establish the force balance, it is convenient to consider the differential energy balance between the electrochemical energy for the addition of volume $dV$ that induced the internal stress to push the Na into the pores, and the counteracting mechanical energy from the dewetting capillary force at the contact circumference that traveled along the pore wall by a distance of $dl$: $dE = I\eta dt = n_p 2\pi r_p \gamma \cos\theta \, dl$. Here, $\eta$ is the reaction overpotential corresponding to the total current $I$ that is responsible for the addition of volume $dV$ during $dt$. $\gamma$ is the surface energy of the in-pore Na metal. $\theta$ is the non-wetting contact angle between Na and pore wall. Solving and eliminating $dl/dt$ from both balance equations yield our simple model for the critical overpotential at the onset of metal penetration, where the number of pores $n_p$ also cancels out:

$$\eta = \frac{2\gamma V_m \cos\theta}{r_p F} \tag{1}$$

Consistent with the results displayed in Fig. 1 and Fig. S1, the critical overpotential is proportional to the reciprocal of the pore size of the separator. The understanding and modeling of this critical process followed the insights from the classic Young-Laplace pressure, which has also been implemented for the gas/water interface in the porous gas diffusion layers of fuel cells.[32,33] However, to the best of our knowledge, our model is the first one handling a solid-liquid interface that directly correlates with the electrochemical driving force.

To validate this simple model quantitatively, experimental data in Fig. 1 were adjusted to match the definition of reaction overpotential in Eq (1). I-R compensations were applied to the critical voltage obtained from the experiments (Table S2, voltages at the points of sudden drops). Despite the range of pore sizes of the separators, Fig. 2b displays a very good agreement between the reaction overpotentials and the reciprocals of pore sizes. The slope yields a surface energy of 2.21 J m$^{-2}$ for the in-pore Na metal in the diglyme electrolyte, which is also close to the surface energy values available in the literature.[34,35] With this validation, our simple model can serve as an important safety tool for battery design and management to set the critical overpotential for the metal anodes based on the chosen separator pore size to prevent the penetrations.

To convert the insights into a more practice-relevant format, the above mathematical relationship can be further extended to link the current density via the Marcus-Hush-Chidsey (MHC) theory,[36–38]

$$k_{red/ox} = k_0 \int_{-\infty}^{+\infty} \exp\left\{-\frac{(x - \lambda/k_B T \pm \eta e/k_B T)^2}{4\lambda/k_B T}\right\} \frac{dx}{1 + \exp(x)} \qquad (2)$$

where $\lambda$ is the reorganization energy, $\eta$ the reaction overpotential, $k_0$ the exchange rate constant, and $k_B$ the Boltzmann constant. The net current density can be calculated by $j = F(k_{red} - k_{ox})$. Plotting the critical reaction overpotentials against the corresponding current densities yields a Tafel plot in Fig. S2. Fitting the data with Eq (2) gives a reorganization energy of 160.15 meV. Note that the current densities are corrected by the porosity of the separators since the ionic flux can only come from the wetted pores of the separator, rather than the total geometric area of the metal anode. With this new fitting result, we can identify a critical boundary for the safe fast charging current densities for different pore sizes (Fig. 2c). In the safe domain, the separator|electrode interface is stabilized by the surface energy that can self-modulate[19] an ideally smooth interface, therefore, no penetration will happen. In the penetration domain, however, the surface energy was interrupted and overridden by the mechano-electrochemical driving force at the pore entrance of the separator, after which the localized reaction and transport dynamics will induce fast penetration and internal shorts.

## 2.3 Confinement- and Current-dependent morphologies

While metal penetrations through porous separators are generally attributed to "dendrite" growths,[8,9,40] it is important to clarify the actual place (confinement) and current density that can induce the dendrites. If only looking at the apparent current density applied and observed in our experiments, transport-limited dendritic growths should not occur. The dendrite initiation requires the concentration depletion near the metal surface at Sand's time, upon the application of an over-limiting current density,[28,41,42] i.e. a current density that is higher than the system-specific limiting current density. Linear sweep voltammetry (LSV) experiments were performed to verify the system-specific limiting current density[43,44] in symmetrical Na|Na sandwich cells with ceramic-coated separators. In the LSV voltammogram, a stabilized transient current after the Ohmic regime indicates the transport limitation in the liquid electrolyte.[43-45]

With only one layer of separator (same as in Fig. 1), the penetration occurred shortly after entering the ohmic region (Fig. S3a). In order to observe the global transport limitation, we increased the number of separators to higher than eight pieces, which indeed exhibit the expected current plateau for calculating the system-specific limiting current (Fig. 3a and Fig. S3b-d).[45] Note that, the mathematical formula of limiting current density allow us to extract the key parameters for the transport kinetics, which will exclude the effect of including multiple layers of separators. Interestingly, unlike the results in other systems[43,44] that only have one current plateau, a second current plateau always emerges (Figs. 3 and S3). To understand this unusual phenomenon, several cells were fabricated and subjected to the same LSV experiment but stopped at the different stages, as labeled in Fig. 3a. The cells were then disassembled for postmortem analysis. It appeared that Na deposits grown before the first current plateau only filled the reservoir created by the washer (Fig. 3b), which was used for confining the geometric area for accurate calculation of the current density. After the first plateau, however, dendrites were initiated as expected (at transport limitation) but unexpectedly grew in the radial direction, and into the gap between the washer and the first separator (Fig. 3b), rather than penetrated through the separator. The lateral growths resulted in an increased active area of the working electrode. Therefore, we hypothesized that the second current plateau is simply *the same limiting current density* but associated with the increased electrode area. The growths in the radial direction within the gap continued until they reached the edge of the separator and short-circuited the cell as reflected by the current spikes. To validate this hypothesis, the same LSV experiment was repeated by using washers with different inner diameters, i.e. different initial deposition areas. Not only the first current plateaus scale with the areas, i.e. yielding consistent current densities, the three cells eventually reached their second current plateaus of similar values (Fig. 3c). Na dendrites observed here showed the typical fractal structures (Fig. 3d). The SEM image of the working electrode displays the transition from the uniform deposits in the reservoir to fractal dendrites near the edge of the electrolyte reservoir defined by the washer (Fig. 3e).

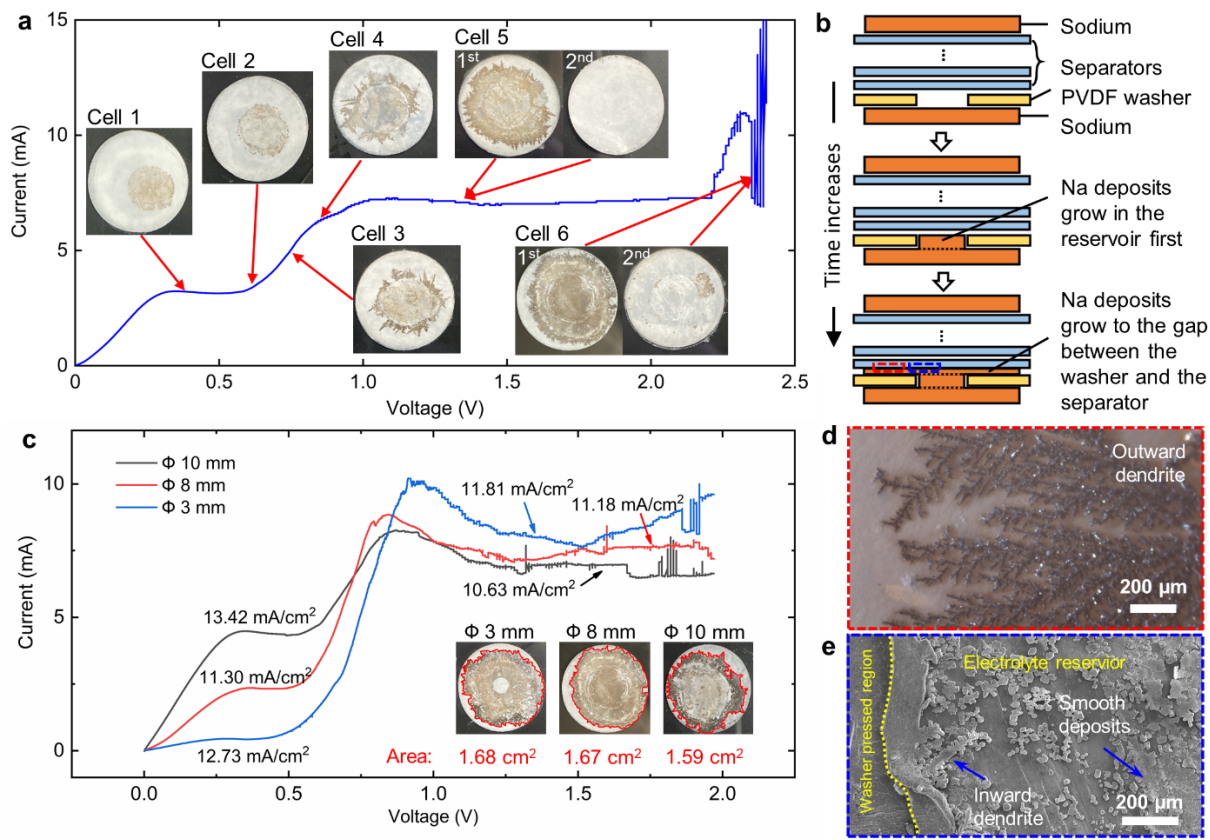

**Fig. 3  Postmortem analysis of the LSV results from sandwich cells.** (a) Two plateaus in the LSV curve. Insets cell 1-6 are the postmortem digital photos of the first separators facing the working electrode from 6 different cells. (b) Schematic explanation of the changes of the electrode working area in the sandwich cell. The PVDF washer is used to define the initial deposition area. (c) LSV results of Na|Na cells with different initial working areas defined by different PVDF washers. Legends indicate the inner diameters of the washers. The current densities associated with the second plateaus were calculated based on the area covered by the sodium dendrites as enclosed by the red outlines on the inset digital photos. (d) Optical digital photo of sodium dendrites on the 1st separator surface covered by the washer. (e) SEM image of the working electrode including the region insulated by the washer and the reaction region. Both dendritic transition near the edge and the smooth deposits before the transition are identified in the reaction region.

Since the limiting current density is inversely proportional to the distance between the two electrodes,[28,42,45] the limiting current density for cells with only one layer of separator should be of order 100 mA cm$^{-2}$, much higher than the penetration current density identified in Fig. 1 and Fig. S3a. In principle, these critical current densities, less than 9% of the limiting current densities, will only generate the ideally stable ingot-type Na growths. Was the penetration really caused by a dendrite that initiated in the big/global confinement? In the case of dendrite growths at current densities higher than the limiting current density (Fig. 3), why would the fractal structure prefer the lateral growth and even avoid entering the separator pores (small/local confinement)? To achieve a complete and self-consistent understanding, *operando* experiments are the best approach to offer the most straightforward answers.

## 2.4 In-pore morphology transition at the transport limitation

As we dicussed in section 2.2 and schematically showed in Fig. 2a, localized growths may enter the pores of the separator at the critical Young-Laplace overpotential, and eventually transition to the dendritic growth at Sand's time.[42] Here, transparent glass capillary cells[42] were taken as the analogue of the penetrating channels to monitor the in-pore penetration behaviors. As a result, three different morphologies were identified under different conditions. At an under-limiting current density, the deposit presents a monolithic piece of ingot with metallic luster and stable transient voltage (Fig. 4a-d), consistent with the morphology stability obtained in the practical coin cells at under-limiting current densities (Fig. 1b,c). At an over-limiting current density, as displayed in Fig. 4e-h and Video S2, a fast-advancing columnar Na dendrite shoots out at the Sand's time. It is worth noting that there could be slow spiky growths (Fig. 4i-l) in electrolytes with a relatively high moisture level,[19] which however cannot be called dendrites as they are not transport-limited growths, but non-ideal reaction-limited growths. Note that, in coin cells tests, these shiny spiky growths could be easily squashed into a flat layer, or washed away during cell disassembly for postmortem imaging, misleading the interpretation of the dynamic interfacial stability. Aside from the fundamental growth mechanisms, the reaction-limited spiky growths are still critical for battery safety since their surface heterogeneity can readily trigger the localized growths into individual pores, followed by in-pore dendrite penetration.

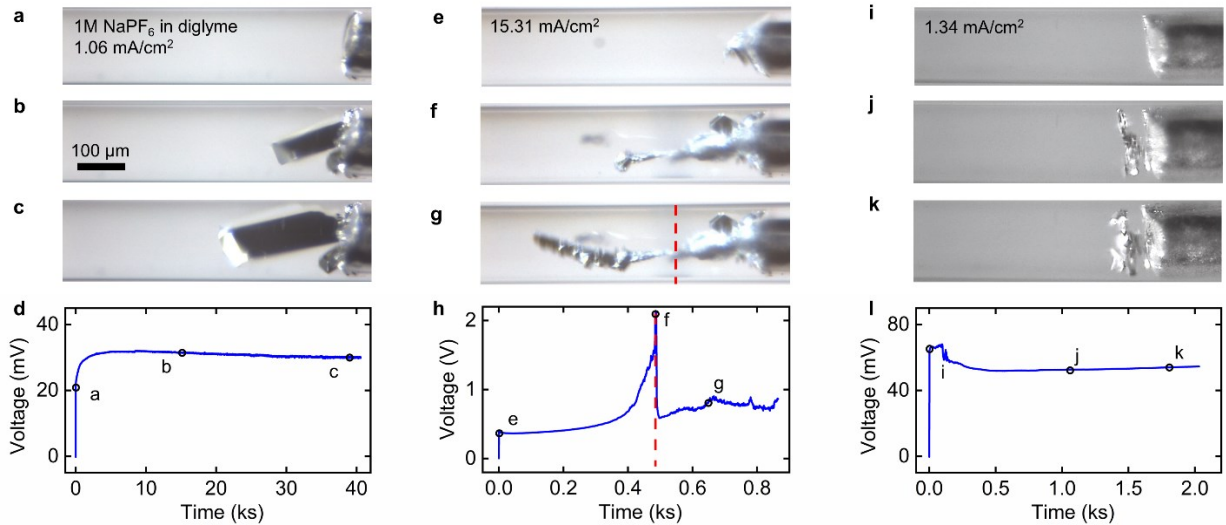

**Fig. 4  Understanding the in-pore penetration dynamic using transparent capillary cells.** (a-d) Snapshots and corresponding voltage response at an under-limiting current density, revealing the ingot-type deposition. (e-h) Snapshots and corresponding voltage response at an over-limiting current density, revealing the shooting of a columnar Na dendrite. Morphology transition occurred at the position and time labeled by the red dashed lines. (i-l) Snapshots and corresponding voltage response at an under-limiting current density with the electrolyte moisture higher than 10 ppm, revealing the spiky deposition.

## 2.5 Orientation preference of Na dendrites

Different from Li metal, Na metal dendrites appear more like the columnar structures observed in alloy solidification,[46] and are apparently different from the sparse lithium dendrites observed in carbonate-based and ether-based electrolytes.[28,42] Columnar metal dendrites are known to have strong orientation preference.[46] As can be seen in Fig. 5a-d, the columnar Na dendrite always grows along the onset direction until touches the inner wall of the capillary. It never modulated itself to grow in the axial direction toward the counter electrode. Every time the columnar Na dendrite bumped to the wall, a voltage spike of over 300 mV was observed (Fig. 5e), which appears to be the driving force for changing the growth direction. This strong orientation preference of Na dendrite growth clearly explained the puzzling phenomenon that Na dendrites in LSV experiments tend to avoid penetrating the pores: it is less energy-expensive for the directional Na dendrite to grow laterally into the gaps, rather than through the tortuous pores of the ceramic-coated separator.

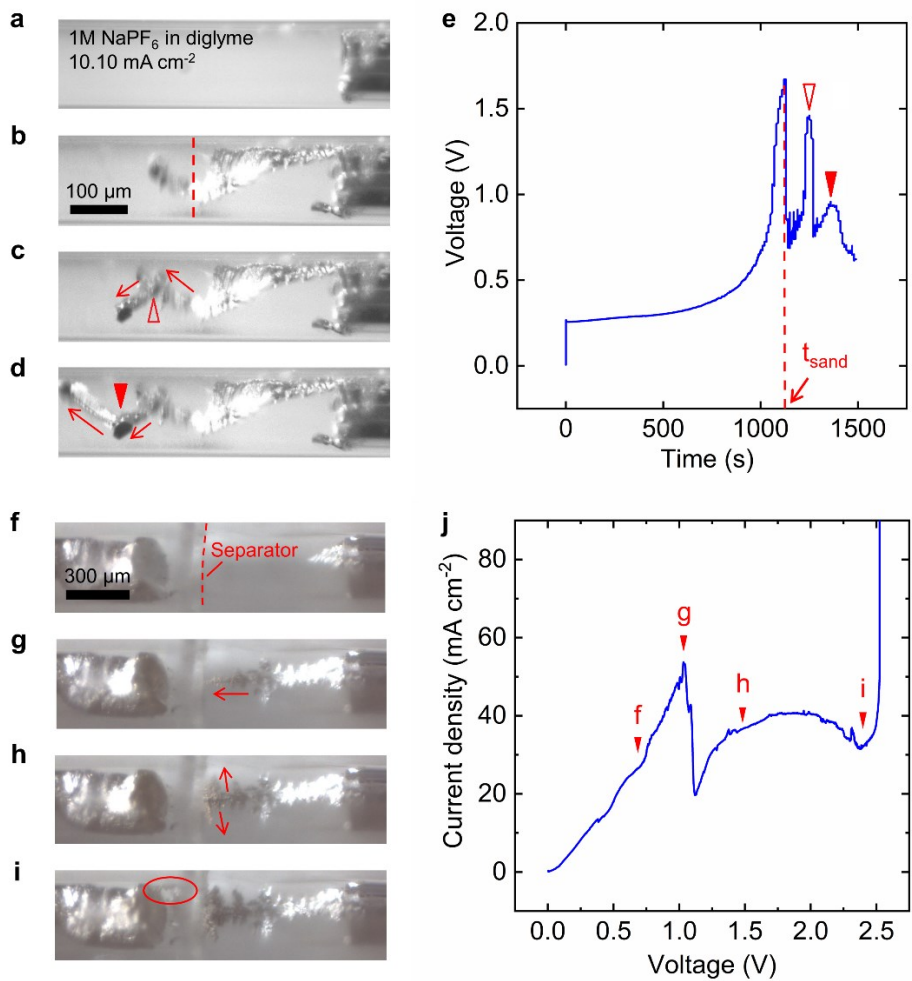

**Fig. 5** *Operando* **validation of the orientation preference of Na dendrites.** (a-e) Snapshots and the corresponded voltage response of Na electroplating at an overlimiting current density with extended time for dendrite growth. The red dashed line indicates the dendrite initiation at Sand's time. Hollow and solid arrowheads are the sites where the dendrite growing pathway is blocked by the glass wall. (f-j) Snapshots of Na dendrite penetrating a layer of separator and the corresponded LSV result. The red arrows show the growing direction of the dendrites and the red circle encloses the Na after penetrating the separator.

This speculation was also confirmed in a junction capillary cell with one piece of trilayer separator. The same LSV test used in Fig. 3 was performed. The initial continuous growth produced an ohmic region in the current-voltage plot (Fig. 5j). The transient current density appeared to flatten out near the limiting current density, accompanied by the emergence of the Na dendrite (Fig. 5f, j). The fast-growing dendrite quickly shortened the distance between the two electrodes, resulting in the increases of transient current density and the intrinsic limiting current density, until reaching the separator. As expected, the dendrite was suddenly blocked by the separator, and started to dwell in the radial direction, consistent with the results in Fig. 3. Note that Fig. 5j is a current-voltage plot, the current dip at time point **g** indicates that the available driving force has to be divided to promote the radial growth, such that the remaining portion of the driving force can only produce a lower current. After a large contact area is formed between the porous separator and the dwelling dendrites, in-pore penetration inevitably occurred through several naturally selected pores at much higher voltages. This extreme case of seeing transport-limited dendrite growth in both the electrolyte reservoir and through the separator pores (Fig. 5f-j) may not happen in practical batteries due to the much smaller inter-electrode distance, i.e. much higher limiting current density. The most concerning situation is that pore-selection (local growth into a pore) and in-pore dendrite growths can happen at relatively low apparent current densities in ideally stable metal anodes. According to our discovery, separators with finer pores or can avoid localized pore-selection dynamics are critical for safe fast charging of metal batteries.

## 3. Discussion

Our results demonstrate that the ideal morphological stability cannot guarantee the absolute battery safety. In the reaction-limited regime, optimizing the pore size of separator, according to our Eq (1) the Young-Laplace overpotential, is a necessary and effective approach to avoid metal penetration even for ideally smooth metal anodes. The critical current density identified for the ideal Na metal anodes is less than 9% of the system-specific limiting current densities, surprisingly lower than that for the non-ideal Li metal anodes (~30%)[27] determined by the same method, which may be attributable to the significantly increased contact area at the separator|electrode interface, where more pore-selection events can be triggered.

Our experiments on concentration polarization in different systems allow rigorous analyses of the limiting current density and Sand's time, from which the two key kinetic parameters, ambipolar diffusivity $D_{amb}$= 2.14 × 10⁻⁶ cm² s⁻¹, and cation transference number $t_+$ = 0.59 (see details in the Supplementary Information) were extracted and cross-validated (Fig. S4). It is worth mentioning that, our experiments of galvanostatic cycling at over-limiting current densities achieved excellent reversibility, without generating dead Na structures that can lead to early failure (Fig. S5 and Videos S3 & S4).

## 4. Conclusion

In this work, we have completed a comprehensive investigation on the fast charging safety limits of ideally stable metal anodes. Our results revealed an alarming discovery that even for ideally smooth Na metal anodes, metal penetration through currently available commercial separators can occur at relatively low apparent current densities. Our analyses of the critical overpotential, however, lead to a simple but predictive Young-Laplace equation that can guide the holistic design of the separator-electrolyte-electrode system. By combining various electrochemical techniques with special transparent cells, our *operando* experiments provided a self-consistent explanation of the penetration mechanisms at multiple scales and in different dynamic regimes. The insights obtained through this work may facilitate the development of better separator|electrode interface that can ensure the safe fast charging of metal batteries.

## Methods

**Materials.** diethylene glycol dimethyl ether (diglyme, anhydrous, 99.5%), sodium cubes (99.9%) were purchased from Sigma-Aldrich. Solvent was dried using molecular sieves for 24 h before use. Sodium hexafluorophosphate ($NaPF_6$, >99%, Alfa Aesar) was purchased from Fisher Scientific. Stainless steel wires, polyvinylidene fluoride (PVDF) sheets were purchased from McMaster-Carr. The glass capillaries for capillary cells were purchased from Narishige Co., Ltd and those for junction cells were purchased from VWR Corporation. Celgard 2325 Polypropylene-Polyethylene-Polypropylene (PP-PE-PP) tri-layer battery separator and ceramic-coated PE battery separator were purchased from MTI Corporation. Whatman 200 nm AAO wafers were purchased from Sigma-Aldrich and 20 nm AAO wafers were purchased from InRedox.

**Cell fabrication and electrochemical testing.** All the cells were assembled in an Ar-filled glove box with $H_2O$ and $O_2$ concentration < 0.5 ppm. For the critical current density experiments, Cu|separator|washer|Na structure was stacked. Electrochemical tests were conducted with a Gamry potentiostat (Reference 600+, Gamry Instruments) and an Arbin battery tester (LBT 20084, Arbin Instruments). The sandwich cells for LSV measurements were constructed in the split test cells purchased from MTI Corporation. Sodium electrodes were cut from the sodium cubes into Φ12 mm and ~ 0.5 mm thick chips. A stack of Na|washer|separators|Na was sealed in the split cell. To obtain better postmortem analysis of the Linear sweep voltammetry (LSV) tests, the ceramic-coated separator was used since the deposits were able to be latched on the separator, rather than on the washer, due to the surface roughness. LSV tests were performed from 0 to 2 V versus $Na^+$/Na at the scan rate of 5 mV $s^{-1}$. *In situ* images were captured by an optical microscope (MU500, AmScope). For capillary cell assembly, the glass capillaries were first pulled 7 mm longer with a vertical type micropipette puller (PC-10, Narishige Co., Ltd) and fixed onto a piece of glass slide using epoxy. Electrolytes were filled in the capillary by the capillary effect from one side. Two pieces of sodium were then pushed to seal the pulled tapering part. Stainless steel wires were used as the current collector to build up the connection with the potentiostat. For junction cell assembly, two segments of capillaries were firstly placed in a row separated by one piece of PP|PE|PP trilayer separator. The connection part was then sealed with PVDF glue and epoxy onto a piece of glass slide. The rest of the assembly was similar to that of the capillary cell.

**Acknowledgement**

This work is supported by a National Science Foundation grant (Award No. 1934122). The materials characterization experiments were partially supported by IMSE (Institute of Materials Science and Engineering) and by a grant from InCEES (International Center for Energy, Environment and Sustainability) at Washington University in Saint Louis. P.B. acknowledges the startup support from Washington University in St. Louis.

**Additional information**

Supplementary information is available online.

**Competing interests**

The authors declare no competing financial interests.